# Confined states in photonic-magnonic crystals with complex unit cell

Yu. S. Dadoenkova,[1,2,3] N. N. Dadoenkova,[1,3] I. L. Lyubchanskii,[3]
J. W. Kłos,[4] and M. Krawczyk[4]

[1]*Ulyanovsk State University, 432000 Ulyanovsk, Russian Federation*
[2]*Novgorod State University, 173003 Veliky Novgorod, Russian Federation*
[3]*Donetsk Physical and Technical Institute of the National Academy of Sciences of Ukraine, 83114 Donetsk, Ukraine*
[4]*Faculty of Physics, Adam Mickiewicz University in Poznań, 61-614 Poznań, Poland*

We have investigated multifunctional periodic structures in which electromagnetic waves and spin waves can be confined in the same areas. Such simultaneous localization of both sorts of excitations can potentially enhance the interaction between electromagnetic waves and spin waves. The system we considered has a form of one dimensional photonic-magnonic crystal with two types of magnetic layers (thicker and thinner ones) separated by sections of the dielectric photonic crystals. We focused on the electromagnetic defect modes localized in the magnetic layers (areas where spin waves can be exited) and decaying in the sections of conventional (nonmagnetic) photonic crystals. We showed how the change of relative thickness of two types of the magnetic layers can influence on the spectrum of spin waves and electromagnetic defect modes, both localized in magnetic parts of the system.

**I. INTRODUCTION**

Study of different types of excitations (electromagnetic, sound and spin waves) in periodic magnetic structures is an object of intensive theoretical and experimental research because of their promising applications in modern photonics [1-3], phononics [4-7] and magnonics [8-11]. Recently, the complex periodic structures composed of multifunctional materials were proposed and the forbidden band gaps of different origin in corresponding frequency regimes were demonstrated. For example, the so called phoxonic crystals where photonic and phononic band gaps coexist in the same structure are discussed in Refs. [12-14]. In our papers [15, 16] we proposed another type of complex superlattices composed of magnetic and dielectric layers which can show both photonic and magnonic band gaps simultaneously. We called these structures as photonic-magnonic crystals (PMCs) in contrast to magnetic photonic crystals [1-3] and magnonic crystals [8-10] where only photonic and only magnonic band gaps exist, respectively.

In Refs. [15, 16] we investigated PMC with simple one-component magnetic periodic structure, *i.e.* equidistant magnetic layers spaced by two-component finite size dielectric photonic crystal (PC). We found essential influence of magnetic subsystem on the photonic spectra and practically unessential changes in magnonic spectra of the PMC due to existence of the dielectric PCs [15, 16]. The principal goal of this paper is to analyze both electromagnetic waves (EMWs) and spin waves (SWs) spectra of the PMC with bi-component magnetic structure which consists of two magnetic layers with different thicknesses in the unit cell. Here we focus here on the confined modes which create opportunity for localization of both types of the excitations in the same part of the structure, which is a necessary condition for EMW and SW interactions, to be explored in future. This idea justifies the detailed studies of the confined excitations (EMWs and SWs) in such complex structures. In the presented paper we perform two separate theoretical studies concerning: (i) the investigations of the photonic ~~defect~~ modes localized in the magnetic layers surrounded by Bragg mirrors in the form of the dielectric PCs and (ii) the calculations of the SWs spectra naturally confined to the magnetic layers separated by dielectric spacers, being the sections of the PC. We show, that the difference in thicknesses of the magnetic layers influences on the spectrum of the photonic modes localized in each type of the magnetic layer. For the appropriate selection of the frequency we can adjust the layers thicknesses to get the simultaneous localization of the electromagnetic waves (EMWs) in both sorts of magnetic layers and to enhance the transmission of the EMWs through the finite PMC. We also show how such structural changes affect the SWs spectrum, which can be additionally tuned by the static magnetic field. Finally, we discuss in general terms the possibility for interactions between EMWs and SWs in the photonic-magnonic structure.

The paper is organized as follows. In Section II we present geometry of the bi-component photonic-magnonic structure. Sections III and IV show results of the numerical calculations of photonic and magnonic spectra, respectively, depending on the magnetic layers thicknesses. In Section V, the Conclusions, we summarize the obtained results and discuss the next problems about interaction of photonic and magnonic excitations in PMC which to be solved.

## II. PHOTONIC-MAGNONIC STRUCTURE

We investigate the 1D periodic structure consisting of magnetic layers of yttrium-iron garnet (YIG) $Y_3Fe_5O_{12}$, separated by non-magnetic dielectric spacers in the form of nonmagnetic dielectric PC composed of alternating layers of titanium oxide ($TiO_2$) and silicon oxide ($SiO_2$), *i.e.*, $(TiO_2/SiO_2)^3 TiO_2$, as shown in Fig. 1.

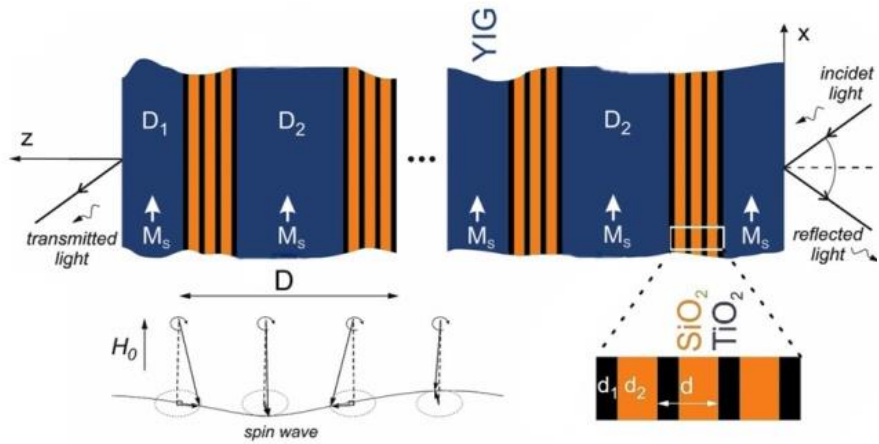

FIG. 1. Schematic of the 1D PMC. The supercell of the period $D = 17.8$ μm consists of two magnetic layers (YIG) of different thicknesses $D_1$ and $D_2$ and two insets of dielectric PC structure $(TiO_2/SiO_2)^3 TiO_2$. The whole structure consists of 7 supercells and an additional YIG layer $D_1$. The YIG layers are magnetically saturated along the in-plane external magnetic field $\mathbf{H}_0$ parallel to the *x*-axis.

The YIG layers are magnetically saturated along the in-plane external field $\mathbf{H}_0$ parallel to the *x*-axis. The thicknesses of $TiO_2$ and $SiO_2$ layers are $d_1$ and $d_2$, respectively, with $d = d_1 + d_2$ being the period of the dielectric PC, and the total thickness of the nonmagnetic spacers is $d_d = 3d + d_1$. The magnetic YIG layers with thicknesses $D_1$ and $D_2$ alternate trough the composite structure. Combination of YIG layers and nonmagnetic composite structure thus forms a double-periodic system or a PMC with the supercell period $D = D_1 + D_2 + 2d_d$. Such a structure can act as a magnonic crystal for the SWs and as a PC for EMWs [15, 16]. The number of the supercells is fixed to 7, and we assume the YIG layers of thickness $D_1$ are located on both left and right sides of the PMC. Thus, the PMC under consideration overall contains 15 magnetic layers and 14 dielectric PC spacers.

For the numerical calculations we choose the thicknesses of the dielectric PCs to be equal $d_d = 1.9$ μm, with TiO$_2$ and SiO$_2$ layers of $d_1 = 0.224$ μm and $d_2 = 0.335$ μm, respectively. We vary the thicknesses of YIG layers, starting with $D_1 = D_2 = D_0 = 7$ μm and keeping total thickness of the supercell fixed as $D = 17.8$ μm. Thus, the thicknesses of YIG layers are $D_1 = D_0 \pm \Delta D$ and $D_2 = D_0 \mp \Delta D$, where $\Delta D > 0$ is the deviation of YIG layer thickness from the nominal value $D_0$. In general, we will consider the PMC, where the first YIG layer thickness is decreasing, $D_1 = D_0 - \Delta D$, thus, it is smaller than the thickness of the next YIG layer $D_2 = D_0 + \Delta D$, which is increasing, i.e., $D_1 < D_2$.

**III. DEFECT MODES IN PHOTONIC BAND GAP**

We consider s- and p-polarized EMWs impinging the right surface of the PMC from vacuum at an incidence angle $\theta$ (with respect to the normal to the interfaces) as illustrated in Fig. 1. The (xz) plane is the incidence plane, and the z-axis is perpendicular to the PMC layers. We use the transfer matrix method to calculate the transmittivity spectra of the finite PMC, as well as to obtain dispersion relation of the corresponding infinite structure. The calculation procedure of the transfer matrix method for the PMC with equivalent magnetic layers is presented in details in [15, 16]. In contrast to the system studied in Refs. [15, 16], in the present paper we deal with the PMC structure with the supercell twice larger, which contains two magnetic layers and two dielectric PC spacers. The transfer matrices for the PMC under consideration corresponding to one supercell has the following form:

$$\hat{T} = \hat{E}_{D1} \hat{T}_d \hat{E}_{D2} \hat{T}_d \quad (1)$$

with $\hat{T}_d$ being the transfer matrix connecting field magnitudes at the boundaries of the magnetic layers separated by one PMC period:

$$\hat{T}_d = \hat{S}_{m1} \hat{E}_{d1} \left(\hat{T}_0\right)^3 \hat{S}_{1m} \quad (2)$$

where the transfer matrix of the dielectric PC subperiod $\hat{T}_0$ is given by Eq. (13) in Ref. [15]. The transfer matrix $\hat{T}_{tot}$ which connects the field magnitudes of the EMW at the right and left surfaces of the PMC can be written as

$$\hat{T}_{tot} = \hat{A}_m \hat{T}^M \hat{E}_{D1} \hat{S}_{m0} \quad (3)$$

The phase incursion matrices $\hat{E}_{D1}$ and $\hat{E}_{D2}$ of the magnetic layers $D_1$ and $D_2$ are

$$\hat{E}_{D1,2} = diag\left[ e^{k_{z,m}^{(+)} D_{1,2}}, e^{-k_{z,m}^{(+)} D_{1,2}}, e^{k_{z,m}^{(-)} D_{1,2}}, e^{-k_{z,m}^{(-)} D_{1,2}} \right] \quad (4)$$

The matrices $S_{m1}$, $S_{1m}$, $S_{m0}$ are given in Refs. [15, 16], and $k_{z,m}^{(\pm)}$ is the z-component of the EMW wave vector in a magnetic layer, where the subscripts ($\pm$) denote two different normal EMWs in the magnetic layers (see Eq. (17) in Ref. [15]). For the sake of simplicity, in this paper we skip the details of analytical description which is given in Refs. [15, 16] and focus on results of the numerical calculations.

We calculated the transmittivity spectra and dispersion relations of the PMC in the wavelength range from 1.3 μm till 3.1 μm, where the dielectric materials $TiO_2$ and $SiO_2$, as well as magnetic YIG are transparent and the damping of the EMWs can be neglected. In numerical analysis we use optical and magneto-optical parameters of the materials ($TiO_2$, $SiO_2$ and YIG) taken from Ref. [15].

First, in Fig. 2(a) we present logarithm of the transmitted light intensity $I_{out}$ normalized by the intensity of the incident *s*-polarized light $I_{in}^{(s)}$ [$\lg T_s = \lg(I_{out}/I_{in}^{(s)})$] as a function of the incidence angle $\theta$ and reduced angular frequency $\omega d/(2\pi c)$, where *c* is speed of light in vacuum, in absence of deviation of YIG layer thickness ($\Delta D = 0$), *i.e.*, when all the magnetic layers thicknesses are equal to $D_0$. The photonic band gap (PBG) of such a PMC contains several equidistant stripes of high transmittivity value (close to unity) which we called inside-PBG modes [15, 16], as they are of the same origin as defect modes of the PCs with defect layers. It has been shown that these modes appear due to presence of the magnetic layers, while the positions of the PBG edges are defined by the composite structure of the nonmagnetic PCs spacers [15, 16].

The behavior of these inside-PBG-modes is different depending on the incident light polarization. The evolution of spectra $\lg(T_s)$ with $\theta$ for different values of YIG layers thickness deviation $\Delta D = 1.5, 3.5$ and $7$ μm are presented in Figs. 2(b), 2(c), and 2(d), respectively. First, the increase of $\Delta D$ leads to increase of the defect modes number [compare Figs. 2(b), 2(c) and 2(d)], but some defect modes are more spectrally wide, then others. For $\Delta D = 1.5$ μm these four "wide" defect modes are almost equidistant. Further increase of $\Delta D$ destroys this equidistance. At the same time, spectrally wider defect modes show tendency to split, as one can see in Figs. 2(c), where $\Delta D = 3.5$ μm, *i.e.*, $D_1 = 3D_2$.

The limiting case of collapse of one of the YIG layers in the supercell, when $\Delta D = 7$ μm, is illustrated in Fig. 2(d) for *s*-polarized light. In this case the PMC consists of the YIG layers of thickness $2D_0$ separated by two nonmagnetic PCs placed together and this new dielectric spacer is the PC with doubled $TiO_2$ layer as a dielectric defect, so that the total thickness of the dielectric spacers is $2d_d$. It is worth to mention that for $\Delta D = 7$ μm the right and left sides of the PMC are the nonmagnetic composite structures, and the total numbers of magnetic layers and nonmagnetic spacers are 8 and 7, respectively, and the latter forms both left and right surfaces of the structure. The defect modes in spectra of such limiting PMCs appear equidistantly in the PBGs, but at its centers the wide defect modes localize, as shown in Fig. 2(d).

The similar results [$\lg T_p = \lg(I_{out}/I_{in}^{(p)})$ vs $\omega d/(2\pi c)$] for *p*-polarized incident light are presented in Figs. 2(e)-2(h). The transmittivity spectra of the PMC demonstrate the same tendencies, as for *s*-polarized light, except the defect modes become wider with increase of the incidence angle. It should be noted that the spectra of *s*- and *p*-polarized light are equal at normal incidence. For all considered PMCs for both *s*- and *p*-polarized incident light, the increase of the incidence angle leads to a shift of the PBG edges and inside-PBG modes to higher frequencies, in the similar way as it was shown in Refs. [14, 15]. Comparing Figs. 2(a)-2(d) and 2(e)-2(f) for the corresponding values of $\Delta D$, one can see that the difference in behavior of the spectra in the cases of *s*- and *p*-polarized input light. The output light intensity $I_{out}$ decreases with increasing of $\theta$ for *s*-polarized incident light, while in the case of *p*-polarized incident light the value of $I_{out}$ increases. We note, that similar dependence was obtained in Ref. [17, 18] for the defect mode of the dielectric PC with complex superconducting defect layer.

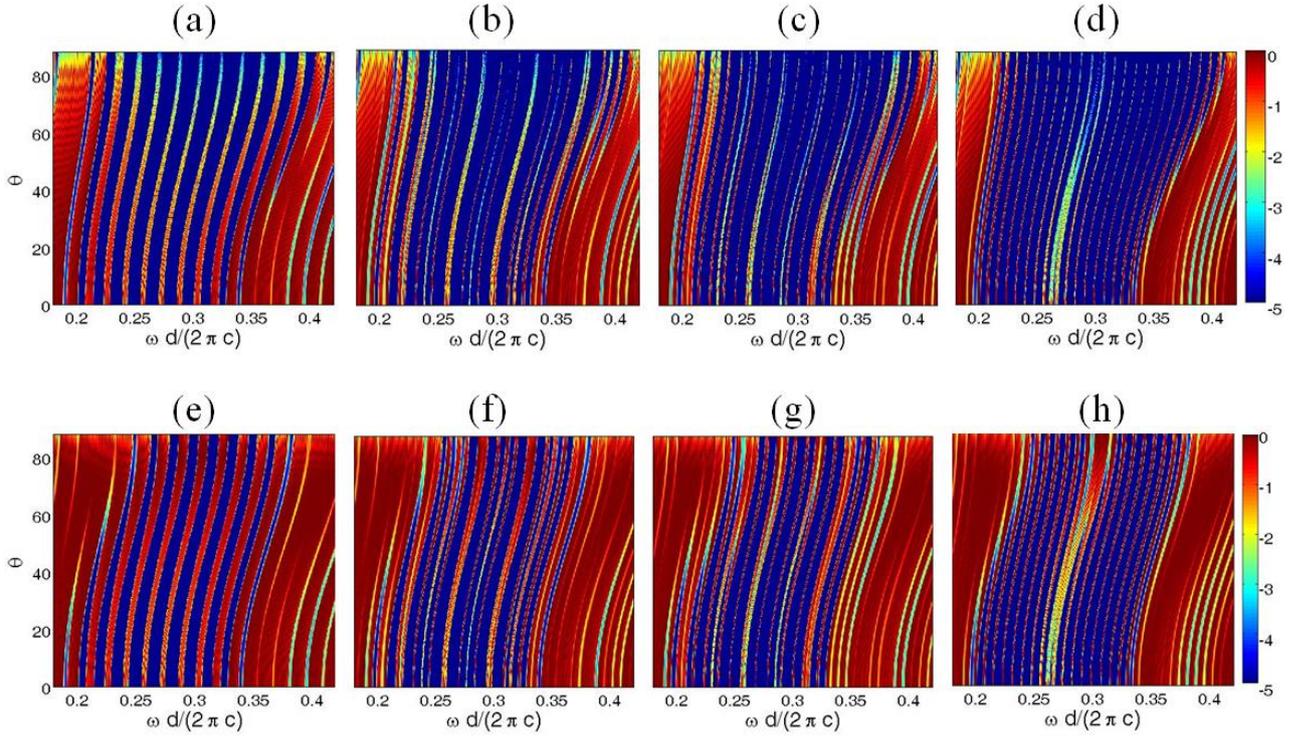

FIG. 2. Transmittivity spectra evolution (in semi-logarithmic scale) with the incidence angle $\theta$ for $s$-polarized (a) – (d) and $p$-polarized (e) – (h) incident light for different values of YIG layers thickness deviation $\Delta D = 0$, (a) and (e); $\Delta D = 1.5$ μm (b) and (f); $\Delta D = 3.5$ μm (c) and (g); and $\Delta D = 7$ μm (d) and (h). The color shows the value of lg $T_{s,p}$.

As it has been mentioned above, the PBGs of the structure under consideration are formed by the dielectric PCs. Figures 3(a) and 3(b) show the transmittivity spectra of $s$- and $p$-polarized EMWs, respectively, for the structure in the limiting case without any magnetic layers, *i.e.*, when $D_1 = D_2 = 0$ and thus the structure represents PC of the structure $[(TiO_2/SiO_2)^3 TiO_2]^{14}$ with 13 defect layers formed by $TiO_2$ layer of the double thickness.

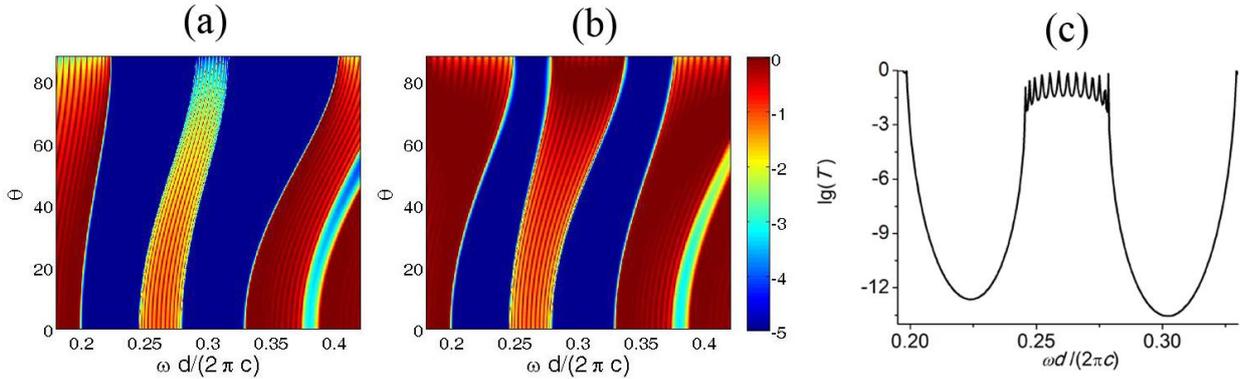

FIG. 3. Transmittivity spectra evolution (in semi-logarithmic scale) with the incidence angle $\theta$ for $s$-polarized (a) and $p$-polarized (b) incident light for the PC without magnetic layers but with the dielectric defects formed by the $TiO_2$ of the double thickness. The color shows the value of lg $T_{s,p}$. (c) Cross-sections of (a) and (b) at normal incidence.

For both polarizations of the incident light, the single wide defect mode is present in the center of the PBG. This mode demonstrates a fine structure, as one can see in Fig. 3(c), and the number of the sub-peaks equals to the number of the double-$TiO_2$ defects. In other words, in purely dielectric PC with periodically distributed defect layers, inside of PBG we obtain the wide transmission peak of the defect mode with fine structure of sub-peaks. This fine structure results from the overlapping of the EMWs localized in the individual defects. As follows from Figs. 2 and 3, the presence of the periodically placed magnetic layers induces the sequence of the PBG defect modes localized inside of the wide YIG layers and cancels the PBG defect modes localized initially in the double TiO$_2$ layers (some remaining of such modes we can also find in the limit $\Delta D \rightarrow 7$ μm, where one type of the magnetic layers disappears [see Figs. 2(d) and 2(h)]. The similar splitting of the transmission peaks for the PBG defect modes localized in YIG is also observed, but it is barely visible due to much larger supercell period $D$ (the supercell contains wide magnetic layers) and weaker overlapping of the EMWs.

We can distinguish yet another possible PMC structure based on the same sub-layers. Apart from the first type PMC discussed above, which begins with thinner magnetic layer ($D_1 < D_2$), we can define also the second type PMC, where the thickness of the first YIG layer is increasing ($D_1 = D_0 + \Delta D$) with respective decrease of the next YIG layer thickness ($D_2 = D_0 - \Delta D$), so that $D_1 > D_2$. The transmission spectra of both types of structures are quite similar. Therefore, in Fig. 2 we present only the results for the PMC beginning with a thinner YIG layer. However, the case when $\Delta D = 7$ μm ($D_1 = 0$ or $D_2 = 0$) requires special attention because of the defect mode localized in double TiO$_2$. The complex structure of the wide peak of this mode (appearing in the center of the PBG) is noticeably different for both types of the PMC structures with $D_1 = 0$ or $D_2 = 0$ as can be seen in Fig. 4.

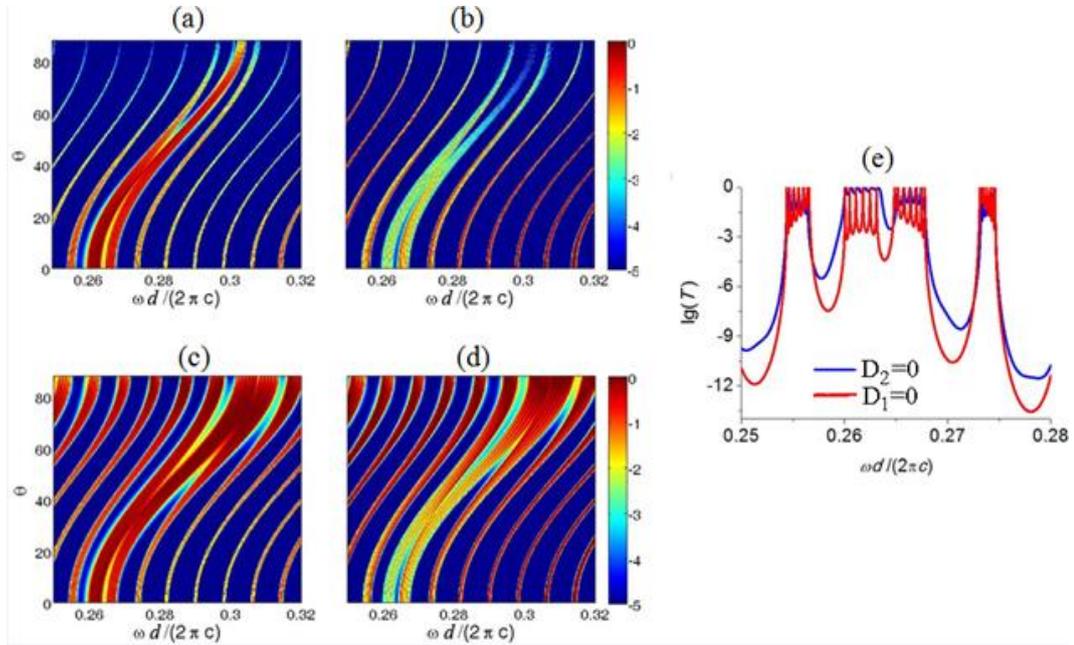

FIG. 4. Central modes in the PBG in the transmittivity spectra (in semi-logarithmic scale) of the PMCs where one of the magnetic layers disappears: $D_2 = 0$, $D_1 = 7$ μm (a) and $D_1 = 0$, $D_2 = 7$ μm (b) for $s$-polarized incident light, and (c), (d), respectively, for $p$-polarized incident light. The color shows the value of lg($T_{s,p}$). (e) Transmittivity spectra in semi-logarithmic scale at normal incidence for the PMCs with $D_1 = 0$, $D_2 = 7$ μm and $D_2 = 0$ and $D_1 = 7$ μm (red and blue lines, respectively).

For $s$-polarized incident light, the central defect mode consists of two branches, which merge in the vicinity of $\theta = 35°$, and split again at the $\theta \approx 45°$, as shown in Figs. 4(a) and 4(b) for the PMCs with $D_1 = 0$ and $D_2 = 0$, respectively. The behavior of these defect modes essentially depends on the structure of the PMC, especially after splitting. The high-frequency branch of the defect mode of PMC with $D_2 = 0$ and $D_1 = 7$ μm remains wide and of high-intensity until grazing incidence, while the corresponding brunch for PMC with $D_1 = 0$ and $D_2 = 7$ μm almost vanishes at high incidence angles [compare Figs. 4(a) and 4(b)]. In the case of $p$-polarized incident light, the branches of the central modes also merge and then slightly split at the same incidence angles, as in the case of $s$-polarized light, but the high-frequency brunch merges with another defect mode at $\theta \approx 65°$

[see Figs. 4(c) and 4(d)]. The splitting of the central mode at low incidence angles is more pronounced in the case of the PMC with $D_2 = 0$, rather than in the PMC with $D_1 = 0$, as follows from Figs. 4(c) and 4(d). At the same time, each of these defect modes has a fine structure, as shown in Fig. 4(e) for normal incidence. The fine structure of the defect modes of the PMC with $D_1 = 0$ is more pronounced than one of PMC with $D_2 = 0$, as follows from comparison of the red and blue lines in Fig. 4(e). In Refs. [15, 16] it has been shown that the fine structure of the inside-PBG modes is caused by the magnetic layers, and the number of the sub-peaks of a single mode is connected with the number of YIG layers. In the system under consideration, the fine structure is more complicated because of mutual influence of the thin and thick magnetic layers. The intensities of the defect modes for any value of the YIG layer thickness deviation $\Delta D$, however, remain larger in the case of the PMC with $D_2 = 0$ rather than in the PMC with $D_1 = 0$, as follows from comparison of Figs. 4(a), 4(c) and 4(b), 4(d).

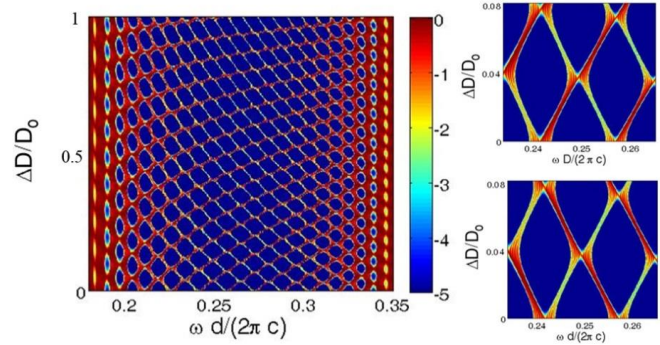

FIG. 5. Evolution of the PBG specter with reduced YIG layers thickness deviation $\Delta D/D_0$ for normal incidence of light. The upper and lower insets show four inside-PBG modes in the transmittivity spectra of PMC beginning with thinner ($D_1 < D_2$) and thicker ($D_1 > D_2$) magnetic layers, respectively, within one period. The color shows the value of $\lg(T)$.

The evolution of the PBG spectrum with reduced deviation of the YIG layers thickness, $\lg(T)$ as a function of $\omega d/(2\pi c)$ and $\Delta D/D_0$, is shown in Fig. 5. The transmittivity demonstrates quasi-periodical dependence on $\Delta D/D_0$, with the period of about $0.08 D_0$, or 0.56 μm, which slightly changes with the frequency. The branches of the defect modes of higher and lower intensities alternate in the opposite order for the PMCs, which begin with thicker and thinner magnetic layers, as one can see comparing the insets in Fig. 5.

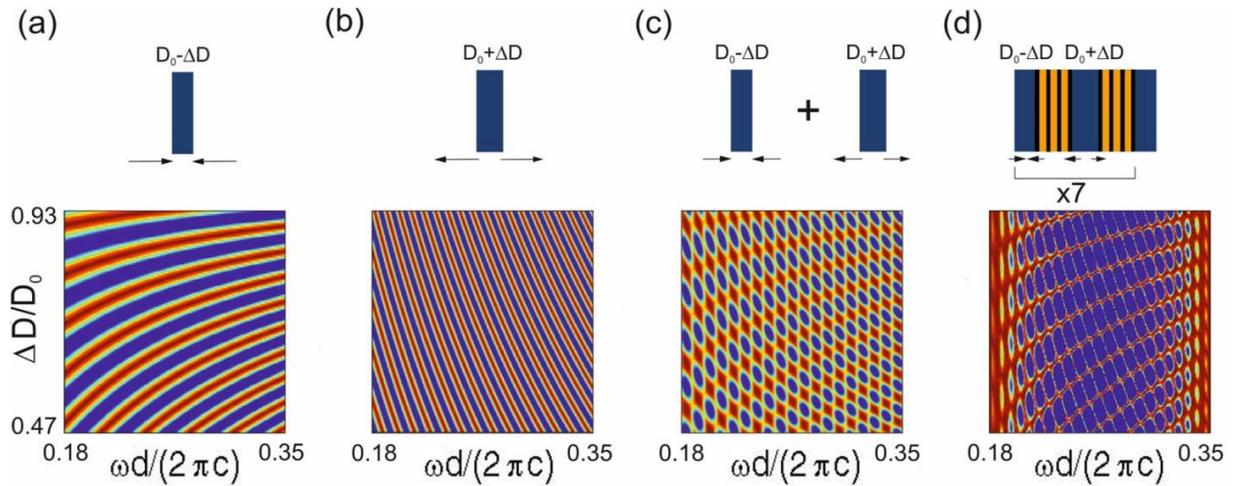

FIG. 6. The transmission of *s*-polarized light thought the single YIG layer as a function of $\Delta D/D_0$ for the normal incidence with (a) decreasing or (b) increasing initial thickness $D_0 = 7$ μm and (c) the contributions of increasing and decreasing thicknesses of two YIG layers. (d) The transmission spectrum of the PMC with 7 superperiods.

In order to explain such dependence of the transmittivity on $\Delta D$, in Figs. 6(a) and 6(b) we plot the transmittivity evolution with $\Delta D/D_0$ for a single magnetic layer with decreasing and increasing thickness, respectively. Decreasing of YIG layer thickness produces a set of lines going up with increase of frequency. On the contrary, increase of $\Delta D$ gives a set of lines going down with the frequency increase. Combinations of these two contributions, which take place in the unit supercell and the PMC, are illustrated in Figs. 6(c) and 6(d), respectively. Thus, there are two types of the defect modes in the PBG of the PMC with two magnetic layers of different thickness in the supercell: one of them is caused by increase of one YIG layer thickness in the supercell, another one is due to its decrease. However, from the transmittivity spectrum of the given PMC it is impossible to define, which inside-PBG mode corresponds to increase or decrease of YIG layer thickness.

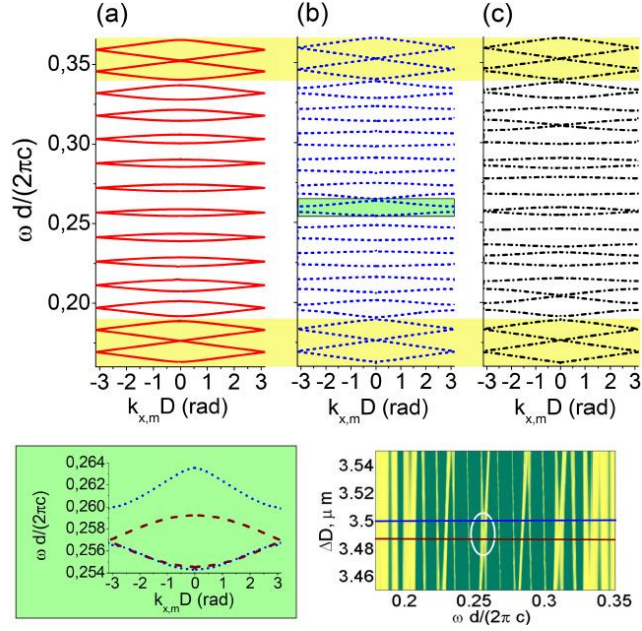

FIG. 7. Transmittivity spectra of the infinite PMC for different values of YIG layers thickness deviation: (a) $\Delta D = 0$ μm; (b) $\Delta D = 3.5$ μm; (c) $\Delta D = 7$ μm. Left inset: central defect modes for $\Delta D = 3.50$ and $3.49$ μm (dotted blue and dashed red lines, respectively). Right inset: defect modes in the vicinity of $\Delta D = 3.5$ μm.

Dispersion of the PMC in the limiting case of infinite number of supercells is illustrated in Figs. 7(a), 7(b), and 7(c) for $\Delta D = 0$, 3.5, and 7 μm, respectively. Here $k_{x,m}$ is the *x*-component of the wave vector of the EMW in a magnetic layer, and yellow areas show the transparency bands of the PMC. In this limit the difference between the PMCs beginning with thinner and thicker magnetic layers vanishes. The inside-PBG modes demonstrate no fine structure, as for an infinite structure the number of YIG layers, which is responsible for splitting of the defect mode, tends to infinity and all sub-peaks merge. Positions of the inside-PBG modes in the spectrum of the infinite PMC remain the same, as in spectra of the finite PMCs beginning with thinner and thicker magnetic layers, as follows from comparison of the left and right insets in Fig. 7. The left inset shows two defect modes in the center of the PMC with $\Delta D = 3.49$ μm and $\Delta D = 3.50$ μm (dashed red and dotted blue lines, respectively). The right inset illustrates the color map of transmittivity spectrum of PMC with $D_1 < D_2$ as a function of $\Delta D$. Here, the intersection of the solid red and solid blue lines with the central defect mode correspond to the similar lines in dispersion of the infinite PMC shown in the left inset. One can see that in both infinite PMC and finite PMC with $D_1 < D_2$ two defect modes merge when $\Delta D = 3.49$ μm, while at $\Delta D = 3.50$ μm these modes already split. Moreover, taking into account Fig. 6 and the right inset in Fig. 7, it is possible to define that the low-frequency brunch of the split defect mode is caused by thickening of YIG layers, and, on the contrary, the high-frequency brunch appears due to shrinking of YIG layers.

**IV. SPIN WAVES IN PHOTONIC-MAGNONIC STRUCTURE**

For SWs, the sections of the dielectric PCs are just homogenous non-magnetic slabs and the whole structure can be perceived as a periodic repetition of thin and thick magnetic layers separated by the spacers of constant thickness. The SWs as the magnetic excitations are confined only in the magnetic part of the structure. Therefore they can coexist in the same spatial areas (magnetic layers) with electromagnetic defect modes discussed in Section III.

The non-magnetic spacers break the exchange interaction between the magnetic layers. The coupling between SWs in successive layers is then ensured by dipolar interaction. For in-plane magnetized films the dynamical demagnetizing field can couple the spin precession in successive layers only if the SWs have the wave vector component which is tangential to the interface [1]. It means that in this geometry and for such magnetic configuration the SWs cannot propagate from layer to layer at normal incidence. The dispersion of SWs (*i.e.*, non-zero group velocity) is observed only for oblique direction of propagation.

The considered system thus can be treated as a sequence of weakly coupled magnetic cavities of two types (being wider and narrower layers), placed alternatively. The change of the thicknesses of these layers allows observing the transition between the single and complex (double) base systems with the period of the structure $D$ remaining unchanged. To calculate the SWs dispersion in this periodic system we used the plane wave method described in Ref. [15]. The method allows to find the spectrum of eigenfrequencies for Landau-Lifshitz equation:

$$\frac{\partial \mathbf{M}(\mathbf{r},t)}{\partial t} = \gamma\mu_0 \mathbf{M}(r,t) \times \mathbf{H}_{\text{eff}}(\mathbf{r},t), \tag{5}$$

which describes the dynamics of the magnetization **M** in effective magnetic field $\mathbf{H}_{\text{eff}}$. Equation (5) was linearized and transformed by Fourier expansion to obtain the eigenvalue problem in the frequency and wave vector space, which was solved for given value of the wave vector $\mathbf{k}_{\text{SW}}$ [15]. In our calculations we fixed the tangential component of the wave vector $k_{\text{SW},x}$ and varied its out-of-plane component in the whole first Brillouin zone. We include both the dynamical exchange and dipolar interactions by adding the appropriate terms to the effective magnetic field [15]. In numerical analysis we use magnetic parameters of YIG taken from Ref. [15].

For $\Delta D = 0$ both types of layers have the same thickness, and the system has actual period equal to $D/2$ — *i.e.*, with one section of the PC and one magnetic layer. The dispersion relation for this structure plotted in the Brillouin zone for period $D$ will be artificially folded at $k_{\text{SW},z} = \pm\pi/D$ [see red dotted lines in Figs. 8(a)-8(c)]. At these points the dispersion curves have self-crossings.

By change of the relative thickness of two types of the magnetic layers we can open magnonic gaps at $k_{\text{SW},z} = \pm\pi/D$ [see the crossings which develop into the first and third gaps marked by G1 and G3 in Figs. 8(b) and 8(c)]. For increasing $\Delta D$ these new gaps (at the frequencies of about 8 and 9.5 GHz) become initially wider [see Fig. 8(d)]. The width of the first gap significantly increases almost monotonically, then starts to saturate, and slightly decreases for the limiting case when one type of the magnetic layer disappears ($\Delta D$ approaches 7 μm and $D_1 = 0$). The width of the third gap significantly decreases for larger $\Delta D$, riches zero and then increases again.

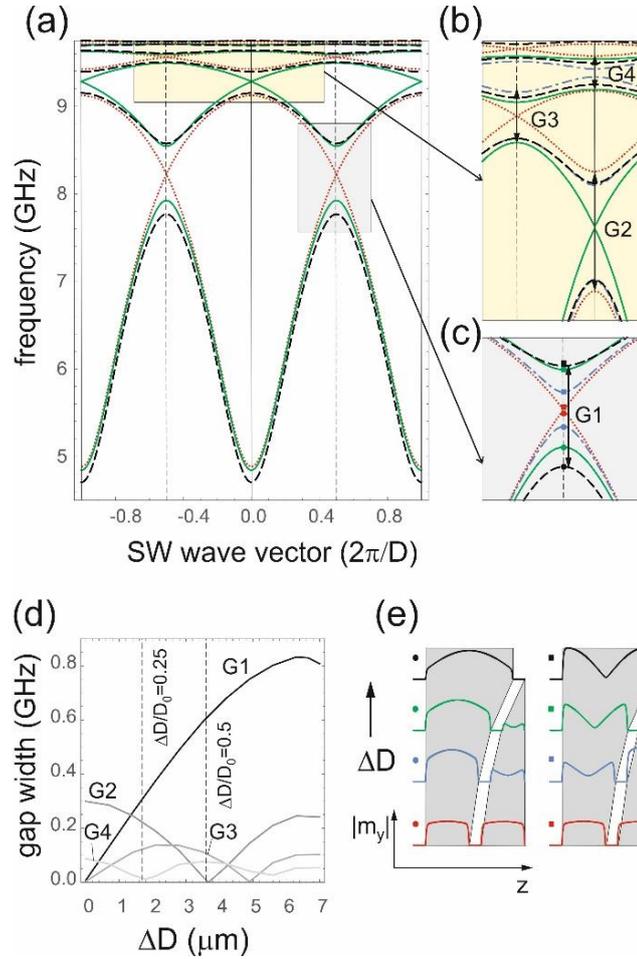

FIG. 8. (a)-(c) Dispersion relation for SW's in considered PMC composed of the sequence of in-plane magnetic layers of thicknesses $D_1 = D_0 - \Delta D$, $D_2 = D_0 + \Delta D$ placed alternatively and separated by non-magnetic spacers (see Fig. 1). The red-dotted, blue-dot-dashed, green-solid, and black-dashed lines correspond to $\Delta D$=0, 1.5, 3.5 and 7 μm, respectively. The magnified insets: (b) and (c) show the regions of four lowest magnonic gaps: G1, G2, G3 and G4. The changes of the width of those gaps in dependence on $\Delta D$ are showed in (d). The amplitude of the in-plane component (*y*-component) of the SWs in a supercell varies in dependence on the relative thickness of both types of the magnetic layers. The profiles of modes above and below the first magnonic gap are shown in (e). The dispersion of SW's propagation in the *z*-direction (from layer to layer) was calculated for fixed value of the in-plane component of wave vector $k_{SW,x} = 0.4$ π/D and for the external magnetic field 10 mT applied in-plane (along the *x*-direction).

The evolution of the width of the magnonic gap already existing for $\Delta D = 0$ can be observed in details in Figs. 8(b) and 8(d). The second and the forth gaps (G2 and G4) start to shrink with the increase of $\Delta D$ and are almost closed for $\Delta D = 3.5$ μm and 1.725 μm, respectively. This corresponds to the case when $D_1$ and $D_2$ are reduced and extended by 50% and 25%, respectively. The further increase of $\Delta D$ makes these gaps wider again.

The interesting issue, which can be important for strength of the magneto-optical interaction, is the distribution of SW's amplitude between the two types of the magnetic layers. We investigated the changes of the SWs profiles at the edges of the first magnonic band in dependence on the difference in thicknesses between the two magnetic layers $\Delta D$. For the lower band [*i.e.*, the first band, marked by circles in Figs. 8(c) and 8(e)] the amplitude is very low for thinner layer. For $\Delta D = 0$ the amplitude equalizes in both layers. By the further change of $\Delta D$ the role of both types of the magnetic layers is swapped. It is also worth to notice, that for thicker layers the dipolar pinning is stronger and the amplitude of SWs close to the interface is lowered in comparison to the amplitude in the center of the layer. For the higher band [*i.e.*, the second band, marked by squares in Figs. 8(c) and 8(e)] the SWs have significantly larger amplitude in the thinner layer. By decreasing $\Delta D$ the amplitude starts to equalize. The further change of $\Delta D$ reverses the role of the layers. In the limiting case of $\Delta D = 7$ μm we have to deal with 1D simple base magnonic crystal, and the mode for the second band has distinctively noticeable zero (node) in the center of the magnetic layer which originates from the minimum gradually deepened with the increase of $\Delta D$.

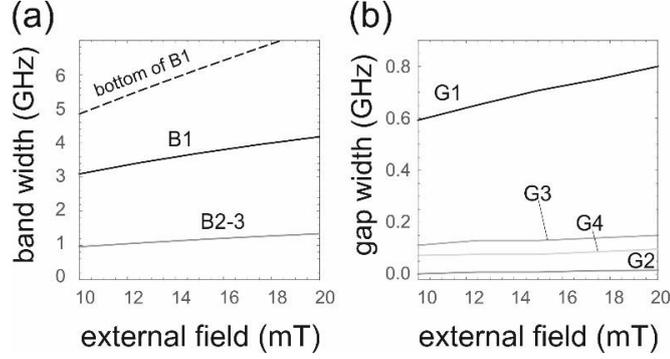

FIG. 9. Widths of the lowest magnonic bands (a) and band gaps (b) for the considered PMC. We fix the difference in thicknesses between the magnetic layers to $\Delta D = 3.5$ μm. The in-plane component of SW wave vector is $k_{SW,x} = 0.4\ \pi/D$. The dashed line in (a) denotes the position of the bottom of the lowest magnonic band.

Interestingly, in magnonic systems we can use the external magnetic field to tune the SW spectrum in the frequency scale. We investigated, how the changes of the external magnetic field magnitude affect the width of the lowest magnonic band gaps and bands. In Fig. 9 we present the results for the structure with the intermediate value of $\Delta D = 3.5$ μm. For this system the second and the third bands merge to one band (denoted here by the symbol B2-3), and the gap between these bands disappear (width of G2 is equal to zero).

We can notice that both the widths of the bands and band gaps are changing almost linearly with the external field increase. By tuning the external magnetic field we both shift and stretch the dispersion relation in the frequency scale. It means that we can affect not only the position of the magnonic bands and gaps but also their width. The change of the width of the bands results in the modification of the group velocity. The widths of the lowest band (B1) and lowest gap (G1) are the most sensitive to the changes of the magnetic field.

## V. CONCLUSIONS

The theoretical studies of electromagnetic waves and spin waves confined in the same areas of the one-dimensional photonic-magnonic crystal are performed. We focused on investigation of influence of the structural parameters, especially thicknesses of the magnetic layers, on the magnonic and photonic excitations spectra in such systems.

The photonic confined modes in the considered photonic-magnonic crystal appear in the YIG layers placed between the sections of $TiO_2/SiO_2$ photonic crystals acting as Bragg mirrors in the frequency ranges corresponding to the photonic gaps of infinite $TiO_2/SiO_2$ stack. The main conclusions referring to the photonic confined modes are:

- we identified two families of the confined modes which can be associated with the magnetic layers of different width;

- for an appropriate selection of the frequency we can adjust thickness of layers to get the simultaneous localization of the electromagnetic waves in both sorts of the magnetic layers (thin and thick) and to enhance the transmission.

The changes of the magnetic layers thicknesses also influence on the spin waves spectra which are naturally confined in the magnetic subsystem. By gradual increase of the thickness of one type of YIG layers and simultaneous decrease the other ones, we observed the opening of new magnonic band gaps at the edges of the first Brillouin zone, whereas some already existing magnonic gaps are about to close for some specific changes of the thicknesses. In the considered system we can adjust the amplitudes of spin waves in the magnetic layers of different type by appropriate choice of their thicknesses. Moreover, we can affect the spin wave spectrum applying an external magnetic field.

The interaction between the electromagnetic and spin waves in the photonic-magnonic crystal can be observed via Brillouin light scattering (BLS) which is very powerful method to study different type of waves including spin wave excitations in multilayered structures [19-23]. Intensity of BLS by spin waves is proportional to the square of Fourier transform of the electric polarization change $\Delta \mathbf{P}(\mathbf{r}, t)$ resulting from the interaction of electromagnetic wave with spin wave [19-22]:

$$I_{BLS} \propto \left|\Delta P(\mathbf{q}, \omega_{inc} \pm \omega_{SW})\right|^2 I^{(0)}(\omega_{inc}) \tag{6}$$

where $I^{(0)}(\omega_{inc})$ is the intensity of incident electromagnetic wave with the frequency $\omega_{inc}$. The symbol $\omega_{SW}$ denotes the frequency of exited or absorbed spin wave. The vector $\mathbf{q} = \mathbf{k}_s - \mathbf{k}_i \pm \mathbf{q}_{SW}$ is the scattering wave vector, with $\mathbf{k}_s$, $\mathbf{k}_i$, and $\mathbf{q}_{SW}$ being the wave vectors of scattered and incident electromagnetic waves, and spin wave, respectively. In linear magneto-optical approximation $\Delta \mathbf{P}(\mathbf{r}, t)$ can be presented in conventional form [19, 21]:

$$\Delta P_\alpha(\mathbf{r},t) = i \sum_{\beta,\gamma} f_{\alpha\beta\gamma} E_\beta^{(0)}(\mathbf{r},t) m_\gamma(\mathbf{r},t) \tag{7}$$

where $i$ is imaginary unit. The indices $\alpha, \beta, \gamma$ go over the Cartesian coordinates $x, y, z$. The symbol $f_{\alpha\beta\gamma}$ represents linear magneto-optical tensor, $\mathbf{E}^{(0)}(\mathbf{r}, t)$ is the electric field of the incident electromagnetic wave, and $\mathbf{m}(\mathbf{r}, t)$ is the dynamical part of the magnetization. If both the amplitude of electromagnetic and spin waves are large in the same regions, then the change of electric polarization due to scattering of electromagnetic wave by spin waves is enhanced [see Eq. (7)] and the intensity of scattered light is increased [24] [see Eq. (6)]. We have achieved this goal in the considered photonic-magnonic crystal by confinement of electromagnetic waves in YIG layers enclosed by Bragg mirrors in the form of fragments of photonic crystals. Repeating the YIG in sequence within the structure of the photonic-magnonic crystal will allow to enhance the signal of the scattered light. Results of the corresponding investigations are in progress and will be published as a separate paper.

Concluding, we showed, that for appropriately designed structures both types of excitations: electromagnetic waves and spin waves can be spatially concentrated in the same areas, which opens the possibility to enhance the interactions between the microwave frequency spin waves and electromagnetic waves from the THz range [25 - 29] This interaction can be potentially useful to control spin wave propagation by electromagnetic waves in magnonics applications or vice-versa in integrated photonics or opto-magnonics devices. In both cases, knowledge of the wave's amplitude distribution inside a structure is crucial for their interaction [30-31].

**ACKNOWLEDGEMENTS**

This research is supported by: the Ministry of Education and Science of the Russian Federation, Project No. 14.Z50.31.0015 (Yu.S.D., N.N.D.); the European Union's Horizon 2020 research and innovation programme under the Marie Skłodowska-Curie grant agreement No 644348 (Yu.S.D., N.N.D., I.L.L. J.W.K., and M.K.); the MPNS COST Action MP1403 "Nanoscale Quantum Optics" (Yu.S.D., N.N.D., I.L.L.); Ukrainian Fund for Fundamental Research under project "Multifunctional Photonic Structures" (I.L.L.).